\documentclass[10pt,letterpaper,twocolumn]{article}
\usepackage{ol}
\usepackage{graphicx}
\usepackage{amsmath,amssymb,textcomp,epsfig,txfonts,subfigure,hyperref}

\date{\today}
\newcommand{\comment}[1]{}

\begin{document}
\twocolumn[
\title{Quantitative modeling of laser speckle imaging}
\author{Pavel Zakharov and Andreas V\"olker}
\affiliation{Department of Physics, University of Fribourg, 1700
Fribourg, Switzerland}
\author{Alfred Buck and Bruno Weber}
\affiliation{Division of Nuclear Medicine, University Hospital
Zurich, 8091 Zurich, Switzerland}
\author{Frank Scheffold}
\affiliation{Department of Physics, University of Fribourg, 1700
Fribourg, Switzerland}
\begin{abstract}

We have analyzed the image formation and dynamic properties in
laser speckle imaging (LSI) both experimentally and with
Monte-Carlo simulation. We show for the case of a liquid inclusion
that the spatial resolution and the signal itself are both
significantly affected by scattering from the turbid environment.
Multiple scattering leads to blurring of the dynamic inhomogeneity
as detected by LSI. The presence of a non-fluctuating component of
scattered light results in the significant increase in the
measured image contrast and complicates the estimation of the
relaxation time. We present a refined processing scheme that
allows a correct estimation of the relaxation time from LSI data. 

\end{abstract}

\maketitle ]
%\linenumbers%
%\section{Introduction}
\comment{ Dynamic light scattering (DLS) is a powerful technique
to study dynamics of the mesoscopic light scattering particles. It
was first introduced as a optical-mixing spectroscopy to study
diffusion in a dilute suspensions of particles
\cite{cummis:broad64}.  Since this pioneering work it has been
widely used in different variations in chemistry, physics, and
biology \cite{berne76}. Two variations which are conceptually the
same are widely used in medicine: laser Doppler velocimetry (LDV)
and laser speckle imaging (LSI)
\cite{briers:review,briers:recon96}. LDV is using the single fast
detector to monitor the fluctuations of scattered light intensity
and estimated the velocity of moving particles from spectrum
broadening due to Doppler effect. Meanwhile }

Laser speckle imaging (LSI) is an efficient and simple method for
full-field monitoring of dynamics in heterogeneous
media\cite{briers:review}. It is widely used in biomedical imaging
of blood flow
\cite{briers:review,dunn:lsi01,weber:imaging,durduran:spat04,dunn:spat05}
since it provides access to physiological processes \textit{in
  vivo} with excellent temporal and spatial resolution.

In a more general context LSI can be considered a simplified
version of the dynamic light scattering (DLS) approach, which
analyzes the temporal intensity fluctuations of scattered laser
light in order to derive the microscopic properties of the scatterers
position and motion. In LSI an image of dynamic heterogeneities is
obtained by analysing \comment{ the spatial statistics of speckles
produced by scattered light and thus allows a full-field
monitoring of dynamics. On of the LSI methods providing the
full-field information is known as a laser speckle contrast
analysis (LASCA) which was introduced as single-exposure
photographic technique\cite{fercher:flow81,briers:retinal82} and
know is widely used as its' digital equivalent
\cite{briers:digital95,dunn:lsi01,weber:imaging}. The original
method is analysing }
 the local
speckle contrast in the image plane. The local contrast $K$ is
defined as the ratio of the standard deviation of the intensity
fluctuations to the mean intensity \cite{briers:review}: $K =
\sigma_{area} / \langle I \rangle_{area}$. In the absence of
scatterers motion the contrast takes a value of $K=1$, while motion
blurs the speckle pattern and thus the contrast decreases. Therefore the
value $K$ can be used to estimate the time scale of intensity
fluctuations and local scatter motion\cite{briers:review}. However
the quantitative interpretation of the LSI data is not
straightforward. Multiple scattering of light can influence the
apparent size of the object due to diffuse blurring. Access to the
local dynamic properties, such as blood flow or Brownian
motion, is complicated by the complex interplay between the
measured contrast and the full fluctuation spectrum of scattered
light. The latter is usually not accessible if technical
simplicity of LSI is to be preserved.

In this letter we present a quantitative approach to analyze LSI
images. We address the problem of spatial resolution and blurring
due to multiple scattering via model experiments and Monte-Carlo
simulations. Further we will demonstrate, that previous attempts to
relate quantitatively LSI images to the microscopic motion have
been hampered by an incorrect data analysis.We present a refined
processing scheme to access this information from a standard LSI
experiment. The main new element of our analysis is to take into
account quantitatively the contribution of the non-fluctuating
part of scattered intensity. We furthermore suggest a simple
experimental procedure that allows to access this important
quantity.

\begin{figure}
\centering
  \includegraphics[width=\linewidth]{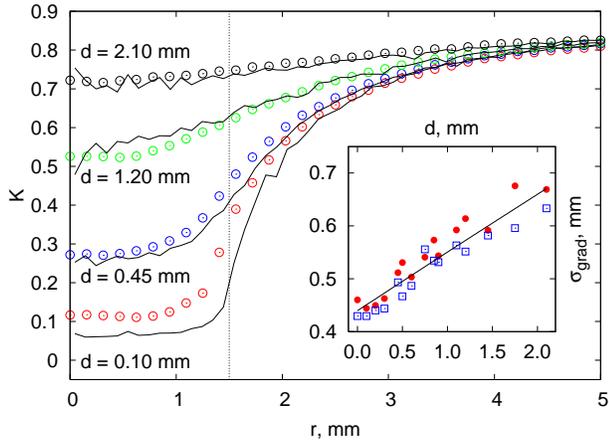}
  \caption{Contrast $K$ as a function of distance $r$ from the center of
  the inclusion for different depths $d$. Curves represent simulation results,  symbols -
  experimental results. Vertical line shows the inclusion
  boundary. Inset: Measured width of the liquid solid-boundary as a function of depth 
  ({\color{red} $\medbullet$} - simulation data, {\color{blue} $\square$} - experiment, 
  line is a guide for the eye).}
  \label{fig:k_vs_r}
\end{figure}

Model experiments have been carried out using a homogeneous block
of solid Teflon and a home-made heterogeneous sample. This medical phantom mimics a liquid
inclusion in solid tissue. It is obtained by milling a cylindrical hole of diameter
 $D = 3$~mm in a block of solid Teflon. A layer of variable thickness $0.45 - 2.1$~mm separates the cylindrical
 inclusion from the interface that is imaged. The void is filled with a dispersion of 710 nm polystyrene particles
 in water. The particle concentration is adjusted to match the optical
 properties of the liquid to the solid (volume fraction ca. $1.3\%$; scattering coefficient $\mu_s = 36$~mm$^{-1}$, mean
free path $l^* \approx 277$ $\mu$m and anisotropy factor $g
\approx 0.9$).  Thus our sample does not show any static
scattering differences. The imaging setup has been described in
detail in Ref.~\onlinecite{voelker:lsi}. Briefly, the cylindric
inclusion with flat base of 3~mm diameter oriented to the surface
is imaged with a CCD camera (PCO Pixelfly, Germany) with a
standard camera objective lens ($f = 50$~mm). The sample is
illuminated with a diode laser (wavelength 785~nm, max. 50~mW).
The beam is expanded by a slow-rotating ground glass in order to reduce
statistical noise\cite{voelker:lsi}. For each depth the contrast
as a function of radial distance $r$ from the inclusion center is
determined. The results of this procedure for different depths of
the inclusion are shown in Fig. \ref{fig:k_vs_r}.

\comment{
Experiment and simulation conditions: temperature 23\textdegree C,
solvent viscosity 0.95 centipoise, particles radius 710 nm, , medium (Teflon) refractive index $n = 1.35$,
, absorption
coefficient $10^-3$~mm$^{-1}$, exposure time $T = 0.130$~seconds,
coherence factor of the setup $\beta \approx 0.867$.
}

For the Monte-Carlo simulation we followed the photon packet
approach of light propagation in a turbid medium \cite{prahl:mc}.
The Henyey-Greenstein phase function was used based on an average
scattering angle $ \langle \cos \theta \rangle$ \cite{prahl:mc}.
The degree of polarization was assumed to decay exponentially\cite{zim:sim00}. The
reflections and refractions on the boundaries were treated
according to Fresnel formulas. As the refractive index of the
Teflon ($n = 1.35$) is very close to water ($n = 1.33$)
interactions with boundaries of dynamic inclusion were not
considered. The photon packets back-reflected from the sample
where registered  within a numerical aperture $0.17$. Only
depolarized photons were registered to match the experimental
conditions. The field auto-correlation function (ACF) $g_1(\tau)$ of the
scattered light was determined from the photon packet history as
explained in Ref.~\citen{durian95:accuracy}.

In order to quantify the effect of image blurring caused by 
multiple scattering we determine the standard deviation of the
contrast gradient $\sigma_{grad}$ . As  can be seen in
Fig.~\ref{fig:k_vs_r} starting from the smallest depths the
apparent width is increased. This increase is of the order of a
few $l^*$, which is the relevant length scale for an incident photon
to propagate laterally. The smearing increases with depth until
for large depths the width becomes of the order of the depth, as
expected for diffuse light propagation \cite{heck97:dark}. Our
results show the importance of diffuse blurring in the image
formation already for small and moderate depths and can thus
provide important guidelines for the analysis of actual
experiments in biomedical imaging
\cite{weber:imaging,durduran:spat04,dunn:spat05}.

\begin{figure}
\centering
  \includegraphics[width=\linewidth]{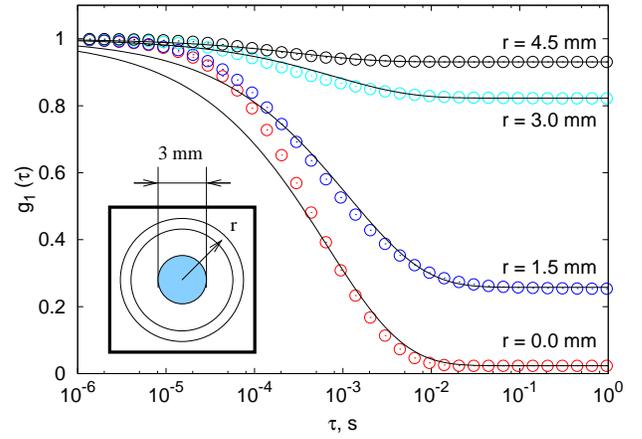}
  \caption{Simulated correlation functions $g_1(\tau)$ for the
  inclusion depth $d$ = 100~$\mu m$. Solid lines show the fits with
  DWS theory with baseline. The relaxation times $\tau_0$ for the fits  were obtained by the contrast $K$ inversion (see Fig.~\ref{fig:inversion}).}
  \label{fig:g1_450}
\end{figure}

A typical set of simulated correlation functions is presented in
the Fig.\ref{fig:g1_450}. If $g_1(\tau)$ is known, then the speckle
contrast $K$ of the time-integrated speckle fluctuations can be
obtained via the following relation
\cite{schatzel:noise90,durian:svs2005}
\begin{equation}
K^2  =  \frac{\langle I^2 \rangle}{\langle I \rangle^2} - 1 =
\frac{2 \beta}{T} \int_0^T |g_1(\tau)|^2 \left ( 1 - \tau / T \right )  d\tau,
%\beta \: |g_1(\tau)|^2 \frac{\Lambda(\tau / T)}{T},
\label{eq:k_vs_T}
\end{equation}
where $T$ is the integration time of the detector and $\beta$ the
coherence factor of the detection optics. It is important to
note that this equation is different to the traditionally used
expression of Fercher and Briers \cite{briers:review}. As
reaffirmed by Durian and coworkers the correct expression, eq.~\eqref{eq:k_vs_T}, has to take into account the effect of
\textit{triangular averaging}
 of the correlation function \cite{schatzel:noise90}.
%\section{Experiment}
As shown in Fig.~\ref{fig:k_vs_r} the resulting values of the
contrast are found in quantitative agreement with the experimental
data without any adjustable parameter.

A comparison of Fig.~\ref{fig:k_vs_r} and Fig.~\ref{fig:g1_450}
immediately reveals the main limitation of LSI. A single contrast
value $K$ is recorded in practice as compared to the complex
correlation function characterizing the full spectrum of intensity
fluctuations and thus particular care has to be taken which
and how information can be extracted.

In our work we have access to both $K$ and $g_1(t)$ and thus we
are able to analyze the LSI detection process quantitatively. We
first analyze the characteristic features of the correlation
functions presented in Fig.~\ref{fig:g1_450}. Overall the
functions are well described by the usual stretched-exponential
form derived for diffusing-wave spectroscopy (DWS) of a colloid
suspension: $g_1(\tau) = \exp ( -\gamma \sqrt{6 \tau / \tau_0})$,
where in our case relaxation time $\tau_0=1/D k_0^2$ characterizes Brownian motion
(diffusion coefficient $D$) and $\gamma \approx 2$ is a
constant\cite{berne76,pine1988}. It is worthwhile to note that for
the case of blood flow the relaxation time $\tau_0$ is inversely
proportional to the root mean square of the flow velocity .

The presence of a non-fluctuating static scattering part, however,
leads to a non-zero baseline. Since the relative amount of the
static light scattering increases with the distance $r$ from
the center, the plateau of the ACF increases as well. It has been noted
previously that a non-zero baseline significantly influences or
even dominates the resulting values of the contrast $K$
\cite{briers:review}. Nevertheless it is rather common in the
biomedical community to convert the obtained contrast values
directly to relaxation times neglecting contributions of static
scattering
\cite{dunn:lsi01,dunn:sim03,dunn:spat05,yuan:optimal05,durduran:spat04,cheng:lsi04}.

If the detected light is composed of dynamic and static
components: $E(t) = E_d(t) + E_s (t)$ the measured field
correlation function is defined as following \cite{boas:spat97}:
\begin{equation}
g_{1} (\tau)  =  \rho \left |g_{\textrm{1 d}} \left ( \tau \right ) \right| + \left (1 -
  \rho \right),
\label{eq:g1_meas}
\end{equation}
where $\rho = \langle I_d \rangle / (\langle I_d \rangle + \langle
I_s \rangle)$ characterizes the fluctuating part of the detected
light intensity and $g_{\textrm{1 d}} \left ( \tau \right )$ is
the field correlation function of the dynamic component. The
resulting contrast value can be calculated by substituting
eq.~\eqref{eq:g1_meas} into eq.~\eqref{eq:k_vs_T}:
\begin{eqnarray}
K_m^2 & = & \frac{2 \beta}{T} \int_0^T \left [ \rho \left |g_{1 d} \left ( \tau \right )
  \right | + \left ( 1 - \rho \right ) \right ]^2 \left ( 1 - \tau / T \right )  d\tau
  \nonumber \\
& = & \beta \left [ \rho^2 K^2_{2\:d} + 2 \rho \left (1 - \rho
  \right ) K_{1\:d}^2 + \left ( 1 - \rho \right )^2 \right ],
\label{eq:k_mixture}
\end{eqnarray}
where $K^2_{\textrm{2 d}} =  1 / T \int_0^T \left | g_{\textrm{1 d}} \left ( \tau \right )
  \right |^2 \left ( 1 - \tau / T \right )  d\tau $  and $K^2_{\textrm{1 d}} = 1 / T \int_0^T
  \left |g_{\textrm{1 d}} \left ( \tau \right ) \right | \left ( 1 -
  \tau / T \right )
  d\tau$ are the normalized variances of the
  intensity and field fluctuations of the dynamic part of the
detected signal with $K_{\textrm{2 d}} <  K_{\textrm{1 d}}$. The
last term in eq.~\eqref{eq:k_mixture} is time independent and
reflects the impact of the static scattering contribution. Thus
the contrast $K_m$ measured in an LSI experiment is given by the
dynamic contrast of intensity fluctuations $K_{\textrm{2 d}}$, the
dynamic contrast of the field fluctuations $K_{\textrm{1 d}}$ and
$\rho$.

\begin{figure}
\centering
\comment{
\mbox{
      \subfigure{
    \includegraphics[width=0.44\linewidth]{pics/contrast_vs_depth.pdf}
} }\mbox{
      \subfigure{
}}
}
     \includegraphics[width=\linewidth]{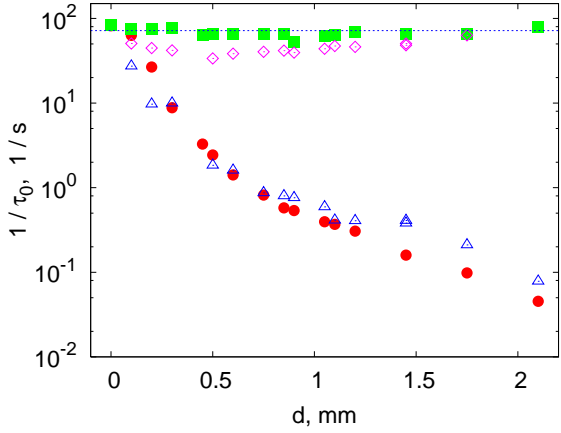}
  \caption{Inverse of the relaxation time $1 / \tau_0$ as a function of
    depth obtained by directly converting the contrast $K$ based on
    Eq.\ref{eq:k_vs_T} neglecting the static part: {\color{red} $\medbullet$} - simulation data, {\color{blue} $\bigtriangleup$} - experiment
    and using the correct procedure based on Eq. \ref{eq:k_mixture}: {\color{green} $\blacksquare$} - simulation and {\color{magenta}
$\Diamond$} - experiment.  Horizontal dashed line - actual value
of the relaxation time. }
  \label{fig:inversion}
\end{figure}

In an actual LSI
experiment an additional processing step has to be introduced in order
to separate the dynamic and static part. The
camera exposure time $T$ is usually larger compared to the
relaxation time $\tau_0$ related to blood flow and subsequent
frames are separated by a period $\Delta t$ larger than $T$.
Since $\Delta t
> T> \tau_0$ two sequential frames are free of the dynamic
component of interest and thus the time and space dependent static
or pseudo-static component can be easily found by
cross-correlating sequential frames  $\rho \left( t, { \bf r}
\right) = 1 - \left [ \langle I \left ( t, { \bf r} \right ) I
\left ( t + \Delta
  t , { \bf r} \right) \rangle_s /
\langle I \left ( t, { \bf r}\right ) \rangle_s^2 - 1 \right
]^{1/2}$.

Fig.~\ref{fig:inversion} shows a comparison of the correct
inversion procedure based on eq. \eqref{eq:k_mixture} as compared
to the one neglecting static scattering \cite{durian:svs2005}.
While the correct procedure produces values comparable to the
relaxation time of Brownian motion $\tau_0$ the direct conversion
can deviate by several orders of magnitude.

In conclusion, we studied the image blurring of an object buried
in turbid media and found that the resolution of the obtained
images can be affected significantly by multiple scattering. We
furthermore introduced a model that reflects the impact of the
static scattering on the quantitative interpretation of LSI
images. A simple procedure has been suggested to perform a
quantitative analysis in actual LSI experiments.

Financial support from the Swiss National Science foundation is
gratefully acknowledged. Correspondence should be addressed to
Frank Scheffold (e-mail, Frank.Scheffold@unifr.ch) or Pavel
Zakharov (e-mail, Pavel.Zakharov@unifr.ch)

\end{document}